\documentclass[iop]{emulateapj}
\usepackage{apjfonts}
\usepackage{graphicx}
\usepackage{color}

\usepackage{epstopdf}
\usepackage{amsmath}

\def\ltsima{$\; \buildrel < \over \sim \;$}
\def\simlt{\lower.5ex\hbox{\ltsima}}
\def\gtsima{$\; \buildrel > \over \sim \;$}
\def\simgt{\lower.5ex\hbox{\gtsima}}

\usepackage{color}

 \newcommand{\tot}{\mathrm{tot}}
\shorttitle{Hierarchical Triple Origin of Compact BH-LMXBs}
\shortauthors{S. Naoz et al}

\begin{document}

\title{Formation of Black Hole Low-Mass X-ray Binaries  in Hierarchical Triple Systems}

\author{Smadar Naoz$^1$, Tassos Fragos$^2$, Aaron Geller$^{3}$, Alexander P.~Stephan$^1$, Frederic A.~Rasio$^{3,4}$ }

\altaffiltext{1}{Department of Physics and Astronomy, University of California, Los Angeles, CA 90095, USA}
\altaffiltext{2}{Geneva Observatory, University of Geneva, Chemin des Maillettes 51, 1290 Sauverny, Switzerland}
\altaffiltext{3}{Center for Interdisciplinary Exploration and Research in Astrophysics (CIERA), Northwestern
University, Evanston, IL 60201, USA}
\altaffiltext{4}{Department of Physics and Astronomy, Northwestern University, Evanston, IL 60208, USA}
\email{snaoz@astro.ucla.edu}

\begin{abstract}
The formation of Black Hole (BH) Low-Mass X-ray Binaries (LMXB) poses a theoretical challenge, as low-mass companions are not expected to survive the common-envelope scenario with the BH progenitor. Here we propose a formation mechanism that skips the common-envelope scenario and relies on triple-body dynamics. We study the  evolution of hierarchical triples, following the secular dynamical evolution up to the octupole-level of approximation, including general relativity, tidal effects and post-main-sequence evolution, such as mass loss, changes to stellar radii and supernovae. 
During the dynamical evolution of the triple system, the ``eccentric Kozai-Lidov'' mechanism can cause large eccentricity excitations in the LMXB progenitor, resulting in three main BH-LMXB formation channels. Here we define BH-LMXB candidates as systems where the inner BH companion star crosses its Roche limit. 
In the ``eccentric'' channel ($\sim 81\%$ of the LMXBs in our simulations), 
the donor star crosses its Roche limit during an extreme eccentricity excitation, while still on a wide orbit.
Second, we find a ``giant'' LMXB channel ($\sim 11\%$), where a system undergoes only moderate eccentricity excitations, but the donor star fills its Roche lobe after evolving toward the giant branch. 
Third, we identify a ``classical'' channel  ($\sim 8\%$), where tidal forces and  magnetic braking
shrink and circularize the orbit to short periods, triggering mass transfer.  
Finally, for the giant channel, we predict an eccentric  ($\sim 0.3-0.6$), preferably inclined ($\sim 40^\circ,\sim 140^\circ$) tertiary, typically on a wide enough orbit ($\sim 10^4$~AU), to potentially become unbound later in the triple evolution. 
\end{abstract}

\keywords{stars: binaries: close, stars: black holes, evolution, kinematics and dynamics, X-rays: binaries}

\maketitle

\section{INTRODUCTION}

Although there is much debate in the literature about the specific formation channel(s) for low-mass X-ray binaries (LMXBs), the ``standard'' scenario 
requires a ``common-envelope'' phase, prior to the compact object formation, in order to produce a tight binary \citep{TvdH2006}.  
This scenario may work for LMXBs containing neutron stars.  However, black hole (BH) LMXBs, which are the focus of this Letter, pose a significant challenge to this paradigm \citep[e.g.][]{PRH2003}.  

LMXBs are abundant in the local universe \citep[e.g.,][]{Fabbiano+06} but their individual properties can only be determined in the Galactic population. To date, the accretors of 
17 Galactic LMXBs have been dynamically confirmed to be BH-LMXBs  \citep{McCR2006}.   
About half (9 out of 17) of the Galactic BH-LMXBs have tight orbits ($\lesssim 18\,\rm hr$) and low-mass ($\lesssim 1\,\rm M_{\odot}$) main-sequence or sub-giant companions \citep{McCR2006}. Mass-transfer from the lower--mass component of a binary to the more massive one leads to the expansion of the orbit, so an additional angular momentum loss mechanism is needed in order for the 9 Galactic compact BH-LMXBs to evolve to their currently observed short periods. This mechanism is traditionally thought to be magnetic braking  \citep[e.g.][]{RVJ1983}, which is believed to operate only in stars less massive than ($\sim 1.5\,\rm M_{\odot}$), as they have a substantial outer convective zone where the dynamo effect can operate and produce the necessary magnetic fields. Hence, the initial companion mass in these systems must have been $\lesssim 1.5\,\rm M_{\odot}$.

Within the standard formation channel explained above, it is unclear how a binary whose primary star is massive enough to form a BH and whose secondary is less massive than $\sim 1.5\,\rm M_{\odot}$ can survive the common-envelope phase without merging \citep{PRH2003}. Considering the energy budget of the system~\citep{Webbink1984} and taking into account all possible sources of energy that can help unbind the envelope of the primary star (e.g., internal energy, recombination energy and enthalpy of the primary's envelope), one finds that a secondary less massive than $\sim1.5\,\rm M_{\odot}$ does not provide enough orbital energy to unbind the envelope of a BH progenitor  \citep{PRH2003,2006MNRAS.366.1415J,Ivanova2013}. 

In this Letter, we propose an alternative formation channel for compact BH-LMXBs that skips the common-envelope phase all together. We show that BH-LMXBs can form due to hierarchical triple-body dynamics, in combination with General Relativity (GR), tidal forces and post-main-sequence stellar evolution. The latter includes mass loss, magnetic breaking, expansion of stellar radii and supernova (SN) dynamics.

It is likely that at least some BH-LMXB progenitors originated in a triple configuration, as most massive stars reside in binaries and higher multiples ($\gtrsim 70\%$ of all OBA spectral type stars \citep{Raghavan+10,Tokovinin14a}). 
From dynamical stability arguments these triple stars must be in hierarchical configurations, where the inner binary is orbited by a third body on a much wider orbit. Interestingly, a substantial number of close binaries with an accreting compact object, mainly LMXBs and their descendants (e.g., millisecond radio pulsars), are known or suspected triples \citep{1988ApJ...334L..25G,1999ApJ...523..763T,2001ASPC..229..117R,2003Sci...301..193S,2001ApJ...563..934C,2007MNRAS.377.1006Z,Tho10,Prodan:2012ey,Prodan+15}. Furthermore, \citet{Iva+10} estimated that the
most efficient formation channel for black hole -- white dwarf X-ray binaries in
dense globular clusters may involve dynamically formed triples.

The study of these hierarchical triples can be done in the secular approximation regime (i.e., phase averaged, long-term evolution). We specifically use the ``eccentric Kozai-Lidov'' (EKL) mechanism \citep{Naoz11,Naoz+11sec}.  In the octupole-level of approximation, the inner orbital eccentricity can reach very high values \citep{Ford00,Naoz+11sec,Naoz+14DM,Li+13,Li+14,Li+15}. 
When coupled with angular-momentum loss mechanisms, like tidal friction, the EKL mechanism can efficiently reduce the orbital separation of the inner binary \citep{Naoz11,Naoz+14stars}, and initiate a mass-transfer phase. Thus, forming a BH-LMXB without the need for a common envelope.

The rest of this Letter is organized as follows: we begin by describing the numerical setup (Section \ref{ICs}), we then describe the results and specifically the LMXB formation scenarios (Section \ref{sec:Formation}) and finally we discuss our results in Section \ref{sec:dis}.

\section{Numerical Setup}\label{ICs}

We study the secular dynamical evolution of triple stars up to  the octupole-level of approximation \citep[e.g.,][]{Naoz+11sec}, including tidal effects \citep[following][]{Hut,1998EKH} and GR \citep[e.g.,][]{Naoz+12GR}  effects for both the inner and outer orbits.  
Tidal interactions operate only on the companion star to the BH, and we adopt a constant viscous time of $50$~yr, similar to \citet{Naoz+14stars}. 
Different viscous time assumptions will result in slight variations in the timescales as well as the relative contribution of each formation scenario.  One of the novel additions here is that we account for the effects of stellar evolution in our calculations. This includes mass loss, the evolution of the stellar radii and the SN of the primary star.  We use the {\tt SSE}   \citep{Hurley+00} stellar evolution code to evolve each individual star.

In all of our simulations, we  require orbital initial conditions that satisfy dynamical stability, such that the hierarchical secular treatment is justified. 
The first condition is long-term stability of the triple, in which we follow the \citet{Mardling+01} criterion:
\begin{equation}\label{eq:Mar} \frac{a_2}{a_1} > 2.8\left(1+\frac{m_3}{m_1+m_2}\right)^{2/5}\frac{ (1+e_2)^{2/5}}{ (1-e_2)^{6/5}} \left(1-\frac{0.3 i_{\tot}}{  180^\circ}\right)      \end{equation}
The second criterion is:
\begin{equation}\label{eq:epsilon}
\epsilon=\frac{a_1}{a_2 }\frac{e_2 }{1-e_2^2}<0.1 \ ,
\end{equation}
where $\epsilon$ measures the relative amplitudes of the octupole and quadrupole terms in the triple's Hamiltonian.  This is numerically similar to the stability criterion, Equation~(\ref{eq:Mar}) \citep[e.g.,][]{Naoz+12GR}.  

We set $m_1=30.8\,M_\odot$, where the mass of the resulting BH is $7\,M_\odot$ \citep{Ozel2010,2015ApJ...800...17F}. We choose $m_2$ from a uniform distribution in the range of $0.8-1.5\, M_\odot$, so the final inner binary can be considered as a LMXB. A more massive tertiary will result in expanding the outer orbit due to mass loss rather quickly,  reducing the strength of the EKL effects. A less massive tertiary will result in suppression of the eccentricity excitation \citep{Tey+13}. Thus, we choose  $m_3$ from a uniform distribution between $0.8$ and $3\, M_\odot$. 

\begin{figure}[!t]
\includegraphics[width=\linewidth ]{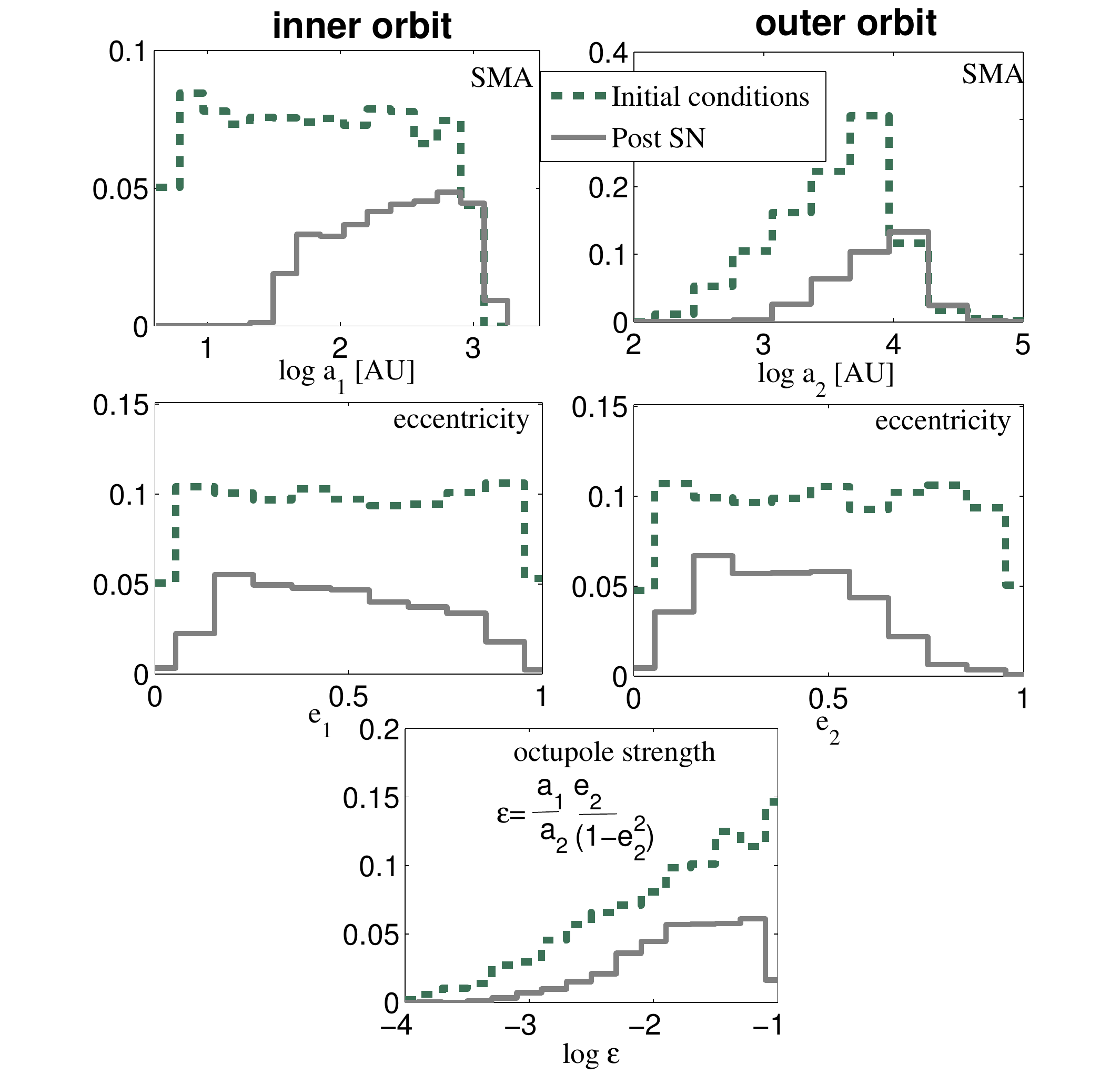}
\caption{The orbital consequences of supernova.
Initial (green dashed lines) and post-SN stable (grey solid lines) distributions of orbital properties (orbital separation and eccentricity) of the inner and outer binary, as well as strength of the octupole approximation. 
About $35\%$  from the initial condition end up in a  post-SN stable hierarchical configuration.} 
   \label{fig:postSN} 
\end{figure}

 We choose the  inner orbit semi-major axis, $a_1$, from a distribution that is uniform in the log between $1000\, R_\odot (\sim 4.7$~AU) and $1000$~AU. The upper limit represents the widest binaries in the Milky Way \citep[e.g.,][]{Kaib+14,Antognini+15col}, while the lower limit is such as to avoid mass transfer before the SN, and  allow EKL precession to be faster than GR.
  We choose the inner (outer) orbit eccentricity, $e_1$ ($e_2$),  from a uniform distribution between $0$ and $1$, following \citet{Raghavan+10}. Our stability requirements (see below), effectively reduce this distribution to be roughly uniform between $0.05$ and $0.85$, see Figure \ref{fig:postSN}.

  We choose the outer orbit semi-major axis, $a_2$,  from a uniform distribution in the log between $a_{2,min}$ and $a_{2,max}=10^4$~AU \citep[e.g.,][]{Kaib+14}. We choose $a_{2,min}$ to be the maximum value from either setting $\epsilon = 0.1$ in Eq.~(\ref{eq:epsilon}), the \citet{Mardling+01} criteria (Eq.~\ref{eq:Mar}), or requiring the quadrupole timescale to be  $\sim6.7$~Myrs, which is the time at which $m_1$ will go SN. This way we guarantee  that the triple configuration will not excite a large eccentricity in the inner orbit before the SN. We should note that we do not take into account here any asymmetries in the SN explosion, as the analysis of the formation history of Galactic BH-LMXBs shows that a SN kick is not required for most of the systems \citep{Willems:2005jf,Fragos2009b,Repetto2012,Repetto:2015vm}. 
  
  We start by following the evolution of the inner binary using {\tt BSE} \citep{Hurley+02}, up to the SN of $m_1$. This allows $m_1$ to lose mass adiabatically   up to  a mass of  $\simlt 9$~M$_\odot$ which causes both orbits to expand. The orbital phase when the star goes SN is chosen  from a uniform distribution and we assume momentum conservation to calculate the new eccentricities (assuming no SN kick).  We find that about $65\%$  of our initial systems either undergo a common envelope prior to the SN or get disrupted by the SN, 
  or simply become unstable according to our criteria in Equations (\ref{eq:Mar}) and (\ref{eq:epsilon}). We then consider the post-SN stable triples   and assume an  isotropic distribution for the initial inclination angle (i.e., uniform in $\cos i$), where we denote the inclination angle of the inner (outer)
orbit with respect to the total angular momentum by $i_1$ ($i_2$), so
that the mutual inclination between the two orbits is
$i=i_1+i_2$. The inner and outer arguments of periapsis   are drawn, respectively, from a uniform distribution. 
  
    In Figure \ref{fig:postSN} we show the results of the SN on the different orbital parameters (specifically on $a_1,a_2,e_1,e_2$ and $\epsilon$). 
The SN tends to widen the orbit and, thus, reduce the strength of the octupole-level of approximation effects. Furthermore, the eccentric orbits are more likely to be disrupted by the SN or become unstable. 




\section{LMXB Candidate Formation Scenarios }\label{sec:Formation}

\begin{figure*}
\includegraphics[width=\linewidth ]{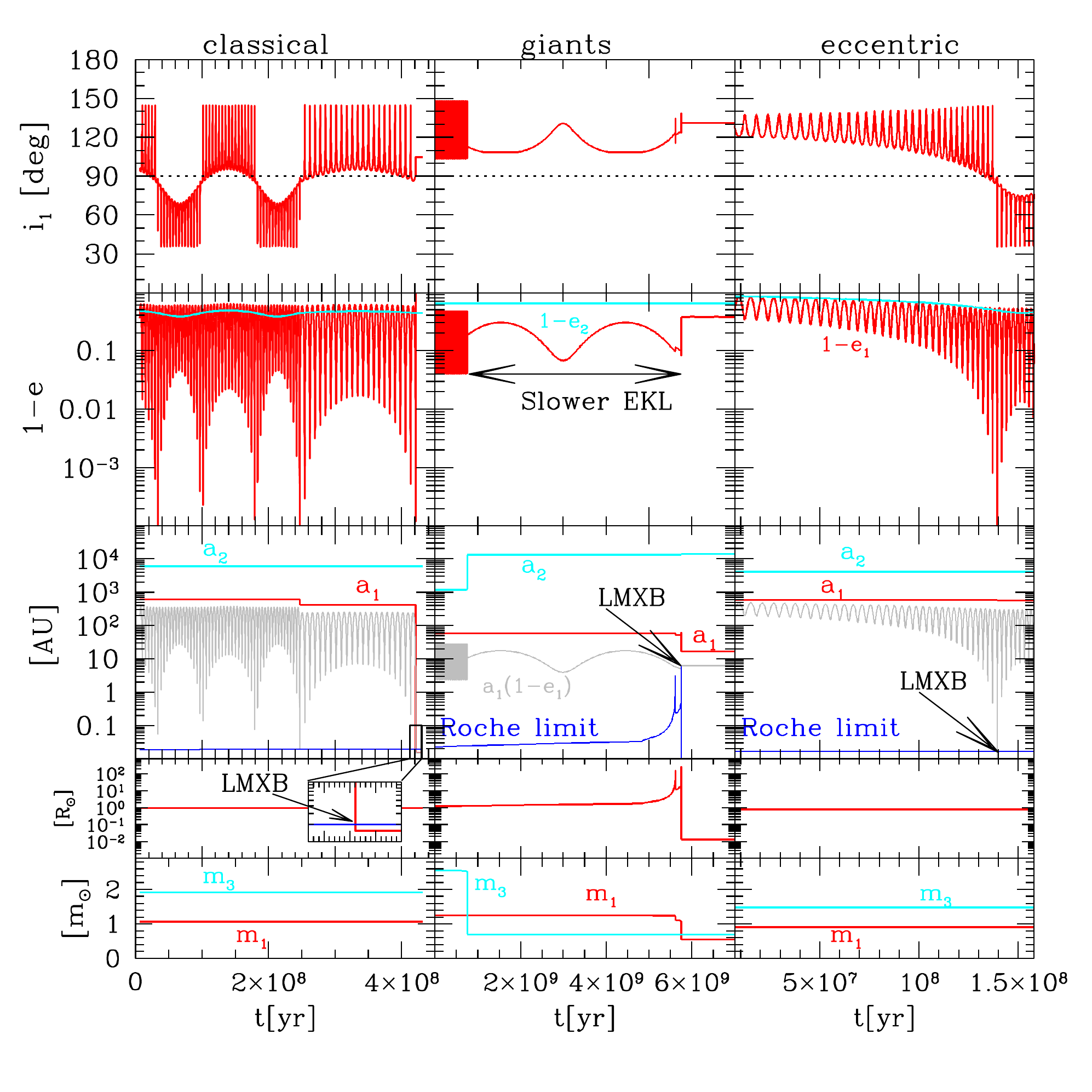}
\caption{Example evolutionary sequences corresponding to the three formation scenarios. We consider from top to bottom the inclination $i_1$, the eccentricity of the inner (red) and outer (cyan) orbits, shown as $1-e$,  the semi-major axis of the inner (red) and outer (cyan) orbits, as well as the pericenter distance of the inner orbit (grey line). Also shown is $a_{\rm Roche}$, as defined in Eq.~(\ref{eq:aRoche}). The last two bottom rows show the BH inner companion star's radius, $R_1$, and mass, $m_1$, as well as the tertiary companion's mass $m_3$. 
{\bf Left panels} show  the time evolution of a system starting right after the BH formation (i.e., post-SN) with $m_1=1.06$~M$_\odot$, $m_2=1.91$~M$_\odot$, $a_1=590$~AU, $a_2=6025$~AU, $e_1=0.64$, $e_2=0.52$, $i=105.3^\circ$, and the argument of pericenter of the inner and outer orbits are set to  $148.8^\circ$ and $332.7^\circ$, respectively. 
The post-SN properties of the system shown in the {\bf middle panels} are $m_1=1.24$~M$_\odot$, $m_2=2.55$~M$_\odot$, $a_1=58.3$~AU, $a_2=1161.1$~AU, $e_1=0.64$, $e_2=0.36$, $i=110.6^\circ$, with the argument of pericenter of the inner and outer orbits set to $338.8^\circ$ and $132.8^\circ$, respectively. Finally in the 
{\bf right panels}  the post-SN properties are $m_1=0.905$~M$_\odot$, $m_2=1.47$~M$_\odot$, $a_1=580.2$~AU, $a_2=4114.2$~AU, $e_1=0.59$, $e_2=0.15$, $i=140.7^\circ$, with the argument of pericenter of the inner and outer orbits set to  $292.4^\circ$ and $186.8^\circ$, respectively.    
For illustration  purposes we allowed the system to evolve beyond the RLO point.}
   \label{fig:ex} 
\end{figure*}

 During the EKL evolution, the inner orbit eccentricity is excited, which can result in crossing of the Roche limit. 
  Following \citet{Eggleton83}, we define the dimensionless    number 
\begin{equation}\label{eq:Roche}
\mu_{\rm Roche}=0.49\frac{ ({m_1} /{m_2})^{2/3} }{0.6 (m_1/m_2)^{2/3}+\ln (1+(m_1/m_2)^{1/3})} \ .
\end{equation}
The Roche limit is then defined by
\begin{equation}\label{eq:aRoche}
a_{\rm Roche}=\frac{R_1}{\mu_{\rm Roche}} \ , 
\end{equation}
where $R_1$ is the radius of the star.
In our simulations, if the BH-companion star  pericenter distance $a_1(1-e_1)$ becomes smaller than $a_{\rm Roche}$ we assume that the star will overflow its Roche lobe. We stop the calculation here and  identify 
 this system as a LMXB candidate.  
%
\begin{figure}
\includegraphics[width=1\linewidth ]{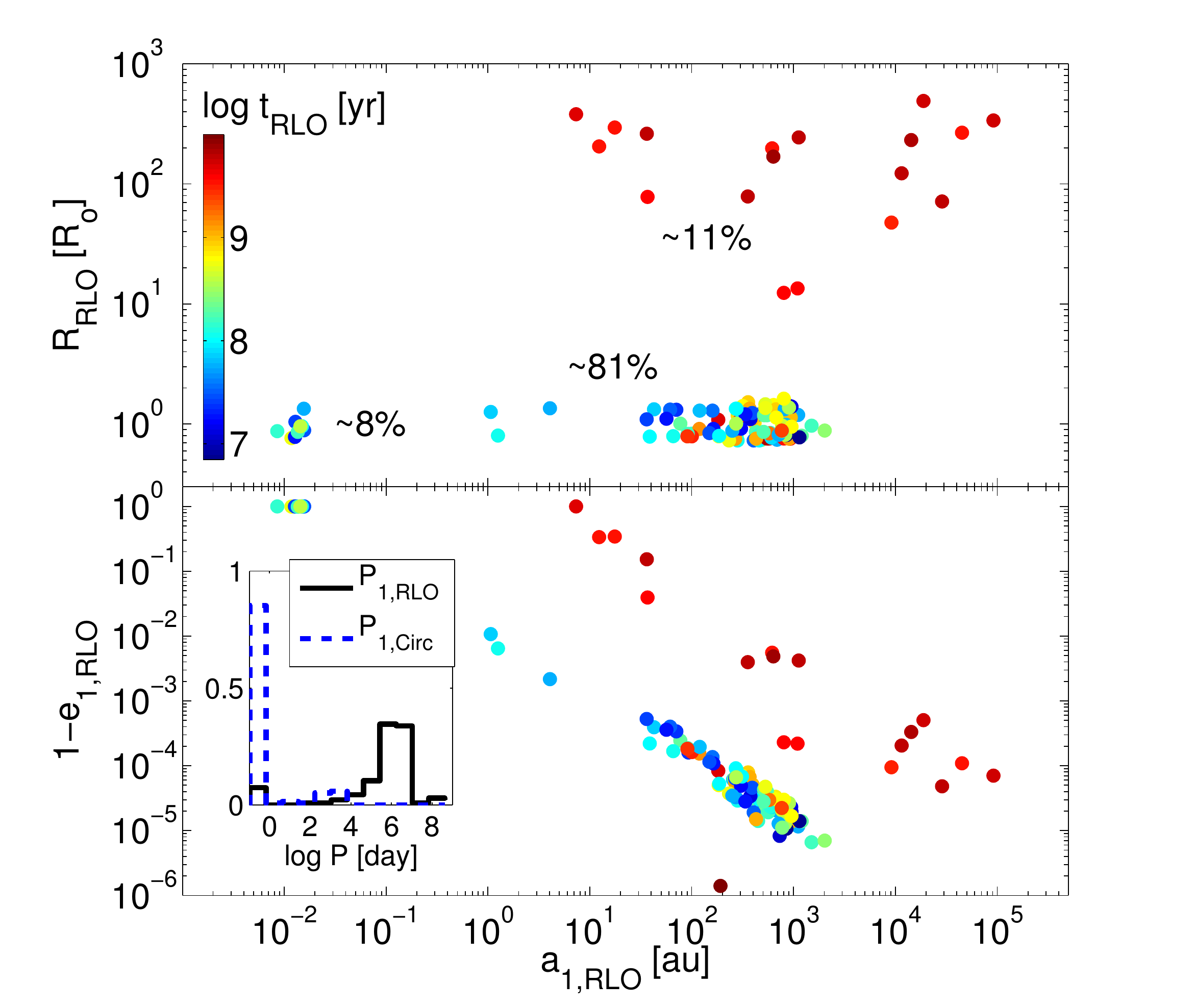}
\caption{Inner orbit parameter distribution of LMXB candidates at the onset of RLO. The three formation channels can be seen clearly in the top panel. The top and bottom panels show the inner companion's radius ($R_{\rm RLO}$) and eccentricity (as $1-e_{1}$), respectively, at the onset of RLO, as functions of the inner orbital separation. The color code is the formation time (in logarithmic scale). 
The inset at the bottom panel shows the period distribution associated with the inner binary at RLO (black line), and the distribution of the circularization orbital period,  assuming that the orbit was instantly circularized at the onset of RLO to an orbit with $a_{circ} = (1-e_{1,\rm RLO})a_{1,\rm RLO}$ (blue dashed line). }
   \label{fig:Innerfinalv2} 
\end{figure}

 
 We denote the parameters of systems that cross their Roche limit with the subscript ``RLO" (Roche Lobe Overflow).  
 In total, we follow the evolution of 2601 post-SN hierarchical triple systems for $10$~Gyr, covering the initial parameter space described in Section~\ref{ICs}. 
  We find a little more than $5\%$ of our runs to be LMXB candidates.  Of these, we identify three formation channels and show an example evolutionary sequence for each one of them in Figure~\ref{fig:ex}. Figure~\ref{fig:Innerfinalv2}  shows the distribution of orbital properties of our population of LMXB candidates, for the inner  orbit, at the onset of RLO. The three formation channels of our LMXB candidates can be easily distinguished in the upper left panel of Figure~\ref{fig:Innerfinalv2}. The three channels are:
 
\begin{itemize}
\item {\bf ``Eccentric":} The majority ($\sim 81\%$) of our LMXB candidates cross their Roche limit when their inner orbit reaches an extremely high eccentricity ($e_{1,RLO} \gtrsim 0.999$). This takes place on a shorter time scale  than the typical extra precession timescale (such as tides, rotational bulge, and the GR precession timescales), and the orbit becomes almost radial. 
 A typical RLO timescale is on the order of $\sim 10^{7-8}\,\rm yr$, while both $m_2$ and $m_3$ 
are on the main sequence (see Figure~\ref{fig:ex}, left panel, and Figure~\ref{fig:Innerfinalv2}).
We note that numerically resolving the maximum eccentricity requires very small time steps. 
Furthermore, due to the chaotic nature of the system, and the limitation of the double averaging method \citep[e.g.,][]{Katz+12,Antognini+13,Antonini+14,Bode+14}, we probably underestimate the maximum eccentricity reached in many of these high peaks. Hence, the overall number of systems that reaches RLO may be higher, and the true distribution of RLO times may be shifted to slightly shorter values than depicted in Figure \ref{fig:Innerfinalv2}. We note that about $\sim 30\%$ of the systems reach such high eccentricities to significantly violate the double averaging procedure \citep[order of magnitude violation of equation (18) in ][]{Antonini+14}. This may cause even larger eccentricities, which can lead to a head-on collision between the star and the BH.  

\item  {\bf ``Giants":} In the second class of systems ($\sim 11\%$), the inner orbit does not reach the same extreme eccentricities for the inner binary to reach RLO. 
The inner BH-companion star eventually evolves to become a giant star, and it is at that  point that it fills its Roche-lobe, generally still at a highly eccentric orbit ($e_1\sim 0.9$). As shown in the example depicted  in the middle panels of Figure~\ref{fig:ex}, after the outer star lost its mass, the eccentricity oscillations slowed down  (labeled ``Slower EKL'' in the Figure), but still continued to be excited to somewhat large values. In the first stellar expansion episode, tides shrank the inner orbit's semi-major axis, which then expanded again due to mass loss in the red giant and AGB phase. 
In the second expansion episode, $a_1$ reduced dramatically, resulting in a ``lucky'' RLO. 

\item {\bf ``Classical":} Finally, we identify a classical class of LMXBs ($\sim 8\%$), where tides play a dominant role. In this case, the eccentricity excitations are large and are gradual enough that tidal and  magnetic braking effects can shrink and circularize the inner orbit at orbital periods of $\lesssim 1$~d.  As depicted in the right panels of Figure~\ref{fig:ex}, the system reached RLO at $\sim$140 Myr, typical of the order 100 Myr timescale after the BH formation inferred from Figure \ref{fig:Innerfinalv2} (for our given choice of viscous time). 
\end{itemize}

\begin{figure}
\includegraphics[width=1\linewidth ]{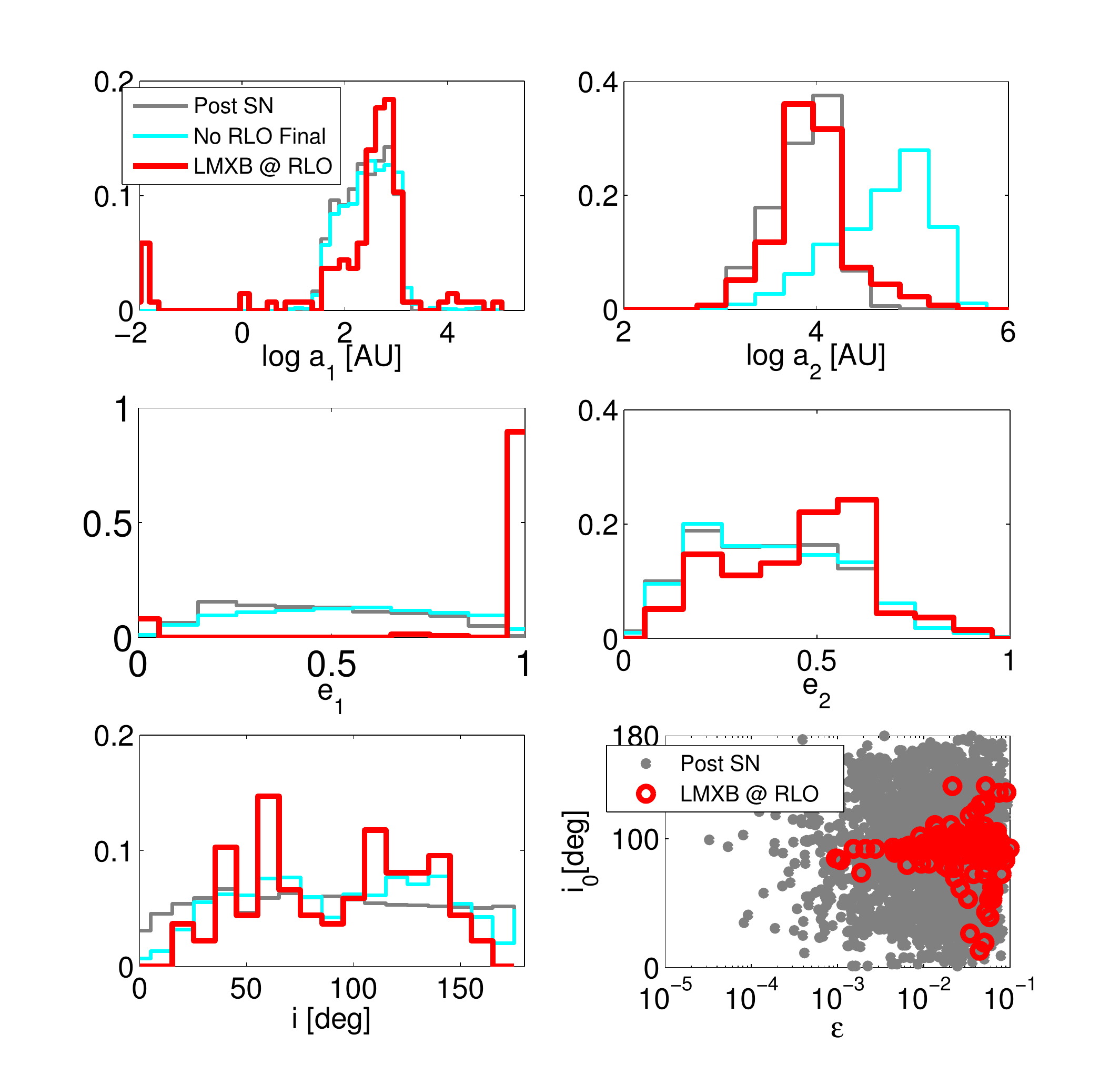}
\caption{Comparison of post-SN orbital property distributions (black) to those of LMXB candidates at the onset of RLO (red), and to systems that completed 10~Gyr of evolution without undergoing RLO (cyan). We consider the following parameters: right panels, from top to bottom show $a_1$, $e_1$, and the mutual inclination $i$; left panels, top to bottom show $a_2$, $e_2$. The bottom right panel shows the post-SN mutual inclination as a function of the   octupole strength ($\epsilon$), where for clarity we omit systems that do not undergo RLO. }
   \label{fig:fOutinal} 
\end{figure}

As expected from the EKL mechanism, in our models the LMXB candidates form preferentially  in systems with initial inclinations between the inner and the outer orbit close to $90^\circ$, and with a larger range of mutual inclination, for stronger octupole contribution (see bottom right panel in Figure \ref{fig:fOutinal}). 
The middle right panel in Figure \ref{fig:fOutinal} shows that modeled LMXB candidates are more likely to be found in triple systems with eccentric outer orbits, which translates to larger octupole strengths [Eq.~(\ref{eq:epsilon})]. 
Note that the initial mutual inclination is not conserved (as depicted in the bottom left panel in Figure \ref{fig:fOutinal}), similarly to  \citet{Naoz+14stars}. 

Finally, BH -- white-dwarf binaries that survive to the end of 10~Gyr of evolution without RLOF are typically on wide orbits (denoted as ``No RLO'' in Figure~\ref{fig:fOutinal}, and with a minimum separation of $8.9$~AU, see top left panel). Their white-dwarf tertiaries are also on a wide orbits (top right panel in Figure~\ref{fig:fOutinal}) and most likely will become unbound due to galactic tides \citep{Kaib+14} or collisions with single stars in the field \citep{Antognini+15col}.

\section{Discussion and Conclusions}\label{sec:dis}

We study the formation of BH-LMXBs through secular three-body interactions. We consider the  octupole-level of approximation of the hierarchical three-body problem, including GR, tidal effects and post main-sequence stellar evolution, which  includes mass loss, inflation of stellar radii  and SN for the BH progenitor.    We run a large Monte-Carlo sample of simulations (2601), of which  about $5\%$ overflow their Roche lobe and become LMXB candidates. We  identify three distinct formation channels for LMXB candidates, which we label as  ``eccentric,''  ``giant'' and ``classical''.

The ``eccentric'' channel describes systems for which their eccentricity is excited to extremely large values (at RLO $e_1\gtrsim 0.999$),  their orbit becomes almost radial and they fill  their Roche lobe. This scenario  represents the majority of our simulated systems ($\sim 81\%$). 
Due to the large eccentricity and the wide inner orbit, the fate of these systems after the onset of RLO is somewhat uncertain. We speculate that  ejecta from non-conservative mass-transfer during periastron RLO will  form a circumbinary disk which may in turn help, through tidal interaction, to quickly circularize the inner orbit.

The ``giants'' formation channel represents a unique combination of EKL and stellar evolution ($\sim 11\%$ from all LMXBs). 
Since the remaining life of the evolved donor star is short, the lifetime of the X-ray bright phase is expected to be short ($\sim$few Myrs),
 and hence the contribution of this scenario to the observed BH-LMXB  population is probably limited.
 

Both the ``eccentric'' and ``giants'' formation scenarios take place during high eccentricity peaks. \citet{Sepinsky2007,Sepinsky2009}  studied the effect of RLO mass-transfer in eccentric  binary orbits, 
assuming a $10^{-9}\rm\,M_\odot \,yr^{-1}$ mass-transfer rate at periastron.
They found that for mass ratios 
we consider here, the circularization timescale due to the mass-transfer is on the order of $1-10\,\rm Gyr$. 
However, their analysis does not include the forced eccentricity oscillations due to a third body or post-main-sequence donors, which we expect will affect the overall dynamics of such systems.

Systems that follow the ``classical'' scenario ($\sim 8\%$ of LMXB candidates in our simulations) form compact LMXBs, with properties similar to the observed ones of Galactic compact BH-LMXBs \citep{McCR2006} and similar to the theoretically estimated properties at the onset of RLO \citep[e.g.][]{Fragos2009b,2015ApJ...800...17F,Repetto:2015vm}. The tertiary companion of such a BH-LMXB is expected to be a white dwarf at a large orbital separation, on the order of several thousand AU, which renders them effectively undetectable. The lifetimes of these systems in the LMXB phase are expected to be on the order of a Gyr, however, their formation timescale is $10^8\,$yr (as deduced from the colors in Figure \ref{fig:Innerfinalv2}), in contrast to $\sim$Gyrs formation timescales predicted by the standard formation channel from primordial binaries \citep{Fragos2013}. The ``classical'' formation channel is the least efficient one in our Monte-Carlo simulations; however, its efficiency is strongly dependent on the assumed strength of the tidal forces, and hence it is difficult to accurately estimate its relative contribution.

Finally, we estimate the current number  of BH-LMXB that may exist in a galaxy such as the Milky Way, as
\begin{eqnarray}
N &\sim & \frac{\Delta N}{\Delta t} \tau_{BH-LMXB} \sim 790  \\
&\times & \left(\frac{\rm SFR}{1~{\rm M}_\odot~{\rm yr}^{-1}}\right) \left(\frac{f_{IMF}}{0.0018~{\rm M}_\odot}\right)  \left(\frac{f_{a_1}}{0.5}\right)  \left(\frac{f_{m_1/m_2}}{0.05}\right)  \left(\frac{f_{KEL}}{0.0175}\right) \nonumber
\end{eqnarray}
for all formation channels combined. The latter number becomes $\sim65$ if we consider only the ``classical'' scenario.
In the above Equation, we  assumed that $100\%$ of all massive stars are in triples, and the inner binaries initially have a flat mass ratio distribution \citep{Sana+12}. The fraction of stars with mass above $30~{\rm M}_\odot$ is  
 drawn from a \citet{Kroupa01} IMF ($f_{IMF}$), $f_{a_1}$ is the fraction of systems with the  inner binary separation assumed in our initial conditions, $f_{m_1/m_2}$ is the fraction of systems with the mass ratio used in our simulations. Finally,  $f_{KEL}$ is the efficiency of EKL mechanism, estimated for all channels to be  $0.35\times 0.05=0.0175$, while for the classical scenario it is  $0.0175\times 0.08=0.0014$. The assumption here is that the lifetime of all BH-LMBX is $\tau_{BH-LMXB} \sim 1$~Gyr, however, this may not be the case for all channels. A proper comparison with the observed population 
 should take into account the transient behavior of LMXBs, as well as the selection effects.
 

%
\acknowledgements
S.N.\ acknowledges partial support from a Sloan Foundation Fellowship. 
T.F.\ acknowledges support from the Ambizione Fellowship of the Swiss National Science Foundation (grant PZ00P2\_148123). A.M.G.\ is funded by a National Science Foundation Astronomy and Astrophysics Postdoctoral Fellowship under Award No.\ AST-1302765.  F.A.R. acknowledges support from NSF Grant AST?1312945 and NASA ATP Grant NNX14AP92G at Northwestern University, and from NSF Grant PHY-1066293 through the Aspen Center for Physics.

 \bibliographystyle{apj}

\end{document}